\documentclass[traditabstract,a4,letter]{aa}  

\usepackage{epsfig}
\usepackage{graphicx}
\usepackage{epstopdf}
\usepackage{color}  
\usepackage{array}   
\usepackage{units}  
\usepackage[breaklinks,colorlinks, citecolor=blue]{hyperref}
\usepackage{natbib}  
\usepackage{multirow}   
\usepackage{amsmath,amssymb}  
\usepackage[english]{babel}
\usepackage[latin1]{inputenc}
\usepackage[T1]{fontenc}
\usepackage{longtable,lscape}
\usepackage{footnote}

\usepackage{txfonts}
\usepackage[toc,page]{appendix} 

\begin{document}


\title{Measuring the hydrostatic mass bias in galaxy clusters by combining Sunyaev-Zel'dovich and CMB lensing
data}
\author{G. Hurier \& R. E. Angulo}

\institute{
Centro de Estudios de F\'isica del Cosmos de Arag\'on (CEFCA), Plaza de San Juan, 1, planta 2, E-44001, Teruel, Spain.\\
\\
\email{ghurier@cefca.es, rangulo@cefca.es} 
}

\abstract{The cosmological parameters prefered by the cosmic microwave
background (CMB) primary anisotropies predict many more galaxy clusters
than those that have been detected via the thermal Sunyaev-Zeldovich (tSZ) effect. This tension has
attracted considerable attention since it could be evidence of physics beyond
the simplest $\Lambda$CDM model. However, an accurate and robust calibration 
of the mass-observable relation for clusters is necessary for the comparison, which
has been proven difficult to obtain so far. Here, we present
new contraints on the mass-pressure relation by combining tSZ and CMB lensing
measurements about optically-selected clusters. 
Consequently, our galaxy cluster sample is independent from the data employed to derive cosmological constrains.
We estimate an average hydrostatic
mass bias of $b = 0.26 \pm 0.07$, with no significant mass nor redshift
evolution. This value greatly reduces the tension between the predictions of
$\Lambda$CDM and the observed abundance of tSZ clusters while being in
agreement with recent estimations from tSZ clustering. On the other
hand, our value for $b$ is higher than the predictions from hydro-dynamical
simulations. This suggests the existence of mechanisms driving large departures
from hydrostatic equilibrium and that are not included in state-of-the-art
simulations, and/or unaccounted systematic errors such as biases in the
cluster catalogue due to the optical selection.}

\keywords{galaxy clusters, CMB, cosmology}

\authorrunning{G. Hurier \& R. E. Angulo}
\titlerunning{Cluster Mass-Pressure relation from tSZ and WL.}

\maketitle
  
\section{Introduction}

Observations of galaxy clusters via the thermal Sunyaev-Zeldovich (tSZ) effect, 
the cosmic microwave background (CMB), and hydrodynamical simulations are in 
tension: adopting the relation between total mass and gas pressure in clusters
predicted by hydrodynamic simulations, the observed abundance of $z<1$ clusters 
is considerably lower than what is expected for the $\Lambda$CDM parameters
preferred by CMB temperature data \citep{planckcmb,planckszc}. 

A possible origin for the tSZ-CMB tension is new physics that would modify the
growth of structure between the last scattering surface and the present day.
However, as shown by \cite{sal17}, the tension can not be solved by simple
$\Lambda$CDM extensions such as massive neutrinos or a time-dependent dark
energy equation of state. An alternative explanation is that there are
deviations from hydrostatic equilibrium much larger than those predicted by
numerical simulations. 

Deviations from hydrostatic equilibrium are commonly quantified via the
hydrostatic mass bias parameter, $b$, defined as the fractional difference
between the true mass of a cluster and that inferred by a gas proxy assuming
hydrostatic equilibrium. Given its importance for cosmology, accurately measuring 
$b$ and understading the relevant astrophysics is one of the primary goals in
the field of galaxy clusters. 

A large number of measurements for $b$ have been performed
\citep[e.g.,][]{med17,ser17,jim17,par17,oka16,bat16,app16,smi16,hoe15,sim15,isr15,van14,don14,gru14,mah13}.
These studies, mostly relying on assumptions of unbiased weak-lensing mass
estimates, obtain $b \simeq 0.20 \pm 0.08$. This value is somewhat high compared
to that predicted by state-of-the-art hydrodynamical simulations \citep[$b \sim
0.1 - 0.2$, ][]{lau13,hahn15,bif16}, but too low to solve the CMB-tSZ tension which
would require $b \geq 0.34$ \citep{sal17}. It is noteworthy that a high value
of $b$ was preferred by \citet{hur17a}, who measured $b = 0.30 \pm 0.07$ using
a joint analysis of tSZ angular power spectrum, bispectrum, and cluster
number-counts. 


The above suggests that systematic errors as well as selection biases could
perhaps be affecting current estimates of $b$, or, conversely, that there could
be physical processes not captured in current hydrodynamical simulations which
would introduce further deviations from hydrostatic equilibrium in clusters.
Solving this issue is crucial to derive reliable cosmological constraints and
potentially detect deviations from the $\Lambda$CDM model.


The recent detection and characterization of the gravitational lensing of CMB
photons \citep{planckphi}, together with all-sky tSZ maps \citep{planckszs},
have enabled independent and robust constrains on $b$. CMB lensing has a
well-determined source redshift distribution ($z \sim 1100$), which reduces the
associated systematic errors. 

The correlation between tSZ and CMB lensing potential, $\phi$, over the
full sky has been measured \citep{hil14}. Unfortunately, the tSZ-$\phi$ angular
power spectrum does not significantly break the degeneracy between cosmological
parameters and $b$. Additionally, it has been shown that this cross-correlation
is significantly contaminated by cosmic infra-red background residuals
\citep[CIB,][]{hur15a}. In a different approach, \citet{mel15} employed the CMB lensing
to measure the individual masses of 61 clusters and derived $b = 0.01
\pm 0.28$. Unfortunately, the low number of systems did not allow these authors
to reach high statistical significance.


In the present analysis, we revisit the measurement of the hydrostatic mass
bias using CMB lensing and tSZ measurements. We consider galaxy clusters
identified by the {\it red sequence Matched-filter Probabilistic
Percolation} ({\it redMaPPer}) algorithm \citep{ryk14} on the SDSS DR8 dataset
\citep{aih11}. We measure the stacked tSZ signal about these clusters binned in
$4$ disjoint richness intervals. We then combine our results with the
CMB-lensing measurements of \citet{gea17}, which estimate the total mass for
the same clusters. With these two measurements, we place
constraints on the cluster hydrostatic mass bias $b = 0.26 \pm 0.07$. 
Finally, we explore potential redshift or mass dependencies and we
discuss the impact of our results on the tSZ-CMB tension.

\begin{figure*}[!h]
\begin{center}
\includegraphics[width=0.7\linewidth]{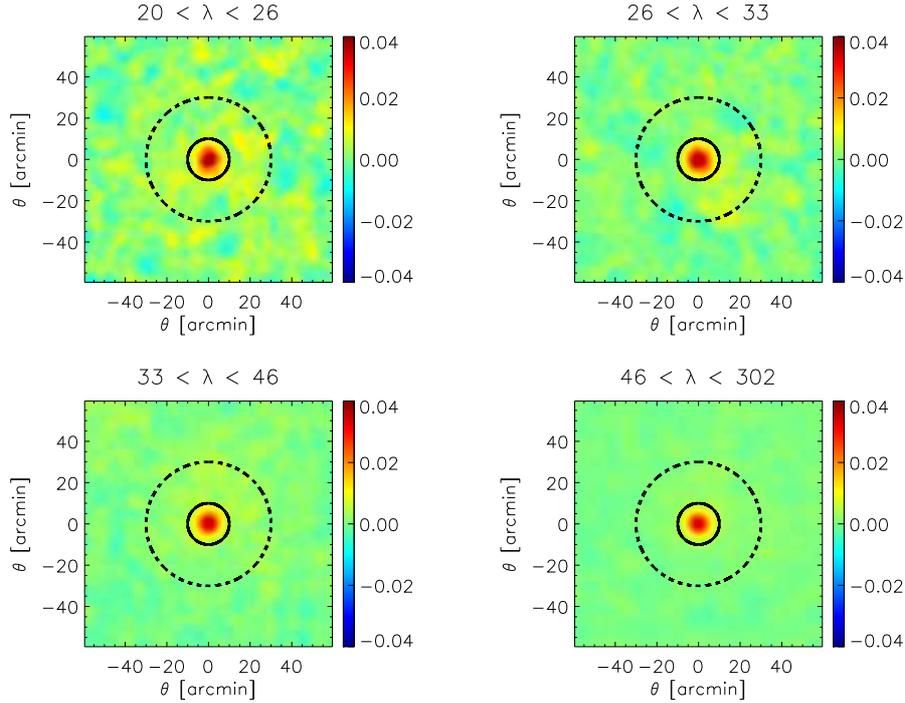}
\caption{Weighted stacking of the Planck tSZ {\tt MILCA} map at 7' FWHM for a
field of view of 2x2 degrees for four richness, $\lambda$, bins. The stacked
tSZ signal is shown in arbitrary units. The solid line black line delimits the
region used to estimated the total tSZ flux, $Y_{\rm TOT}$ and the dashed black
line delimits the region used to estimate the zero-level of the tSZ stack
maps.} \label{sred}
\end{center}
\end{figure*}

\section{Stacked $y$ signals about redMaPPer clusters}
\label{stack}

The core of our analysis is to combine the recent CMB weak lensing measurements
of \citet{gea17} with suitable tSZ estimates for the gas pressure in clusters.
The CMB weak lensing convergence field is given by $ \kappa = \Sigma
(R)/\Sigma_{\rm crit}, $ where $\Sigma (R)$ is projected mass density and
$\Sigma_{\rm crit}$ is the critical mass density,

\begin{align}
\Sigma_{\rm crit} = \frac{c^2}{4 \pi G} \frac{D_{\rm OS}}{D_{\rm OL}D_{\rm LS}},
\end{align}

\noindent with $D_{\rm OS}$, $D_{\rm OL}$, and $D_{\rm LS}$ the angular
diameter distance between the source and the observer, the lens and the
observer, and the source and the lens, respectively. For CMB weak lensing, the
source is the last scattering surface and the lens is the cluster.
\citet{gea17} used convergence maps built by the Planck satellite to estimate
the average total mass, $M_{200}$, of clusters in the SDSS DR8 {\tt redMaPPer} 
clusters catalogue \citep{ryk14} in four richness bins. We now describe our
procedure to estimate the gas pressure for the same clusters.

The intensity of the tSZ signal, $y$, is proportional to the electronic
pressure, $P_e$, of the intra-cluster medium integrated along the line-of-sight
$l$ \citep{sun72},

\begin{align}
y = \frac{k_{\rm B}\sigma_{\rm t}}{m_{e}c^2}\int P_e {\rm d}l .
\end{align}

\noindent where $k_{B}$ is the Boltzmann's constant, $\sigma_{t}$ is the
Thomson's cross section, $m_e$ is the electron rest mass, and $c$ the speed of
light. Thus, $y$ maps can be used to estimate $P_e$ and the hydrostatic mass
bias when combined with independent cluster mass estimates. 

Here, we employ the {\tt MILCA} $y$ full-sky map \citep{hur13, planckszs} at 7'
FWHM resolution \citep[previously used in][]{hur15b} in a field of view of 
$2\times2$ degrees about each cluster. It is interesting to note that $y$ maps constructed
through component separation from multi-frequency data, e.g. {\tt MILCA},
are biased tracers of the pressure since they neglect relativistic
corrections \citep{wri79,hur16}. This induces a significant bias on $y$ 
\citep{hur17b}, but it can be corrected for as we will describe later.

For clusters with significant contamination from radio sources, we
re-compute the {\tt MILCA} map adding extra spectral constraints to reduce the
contribution from radio sources \citep[see][for more details]{hur13}. This
procedure significantly increases the noise level, thus, we apply it only 
to clusters with a clear contamination, as identified in Planck's 70 GHz map.

It is well known that Planck tSZ maps suffer from infra-red emission
contamination, especially from CIB \citep{pug96}. This contamination is
particularly important for the tSZ-CMB weak lensing cross correlation power
spectrum \citep{hur15a}. However, most of this emission originates at high
redshift and, by considering the CIB-leakage transfer function in the tSZ {\tt
MILCA} map \citep{planckszcib2}, this contamination can be neglected for the
stack of our clusters (which are at $z \leq 0.6$).

Following \citet{gea17}, we now stack the tSZ signal about {\tt redMaPPer} clusters
split into four richness, $\lambda$, bins: [20,26[, [26,33[, [33,46[, and
[46,302]. We note that the tSZ effect and the lensing have different
dependencies with the cluster mass and redshift. The lensing
signal scales as $M_{500}/\Sigma_{\rm crit}$, where $M_{X}$ is the mass
contained inside $R_{X}$ (the radius of a sphere with an average density equal
to $X$ times the critical density of the Universe). Whereas, the tSZ flux
within a $R_{500}$ aperture, $Y_{500}$, varies as $M^{1.79}_{500} E(z)^{2/3}$
\citep{planckszc}.

Therefore, we have weighted the tSZ signal associated to each cluster so that
it contributes in the same way as it does to the CMB lensing. Specifically,
within a given richness bin we constructed the stacked map, $Y(\vec{\theta})$, as:

\begin{equation}
Y(\vec{\theta}) = \frac{\sum_i w_i y_i}{\sum_i w_i}, \, \, \, w_i  = \frac{1}{\lambda_i^{0.79} E(z_i)^{2/3} \Sigma_{\rm crit}(z_i)}, 
\end{equation}

\noindent where $y_i$ is the tSZ {\tt MILCA} map about the $i$-th cluster
in the richness bin considered. Note that we have assumed that $M_{500} \propto
\lambda$. Note also that, if the above dependences are neglected, a combined 
analysis would have led to significantly biased results ($\simeq$50\% on $b$ for the
[46,302] richness bin).

The final tSZ stacked maps are presented on Fig.~\ref{sred}. Overall, the tSZ
signal is detected at 9 and 30 $\sigma$ for the lowest and highest richness
bins, respectively.


\section{Measurement of the hydrostatic mass bias}
\label{bias} 

\begin{figure}[!h]
\begin{center}
\includegraphics[width=0.8\linewidth]{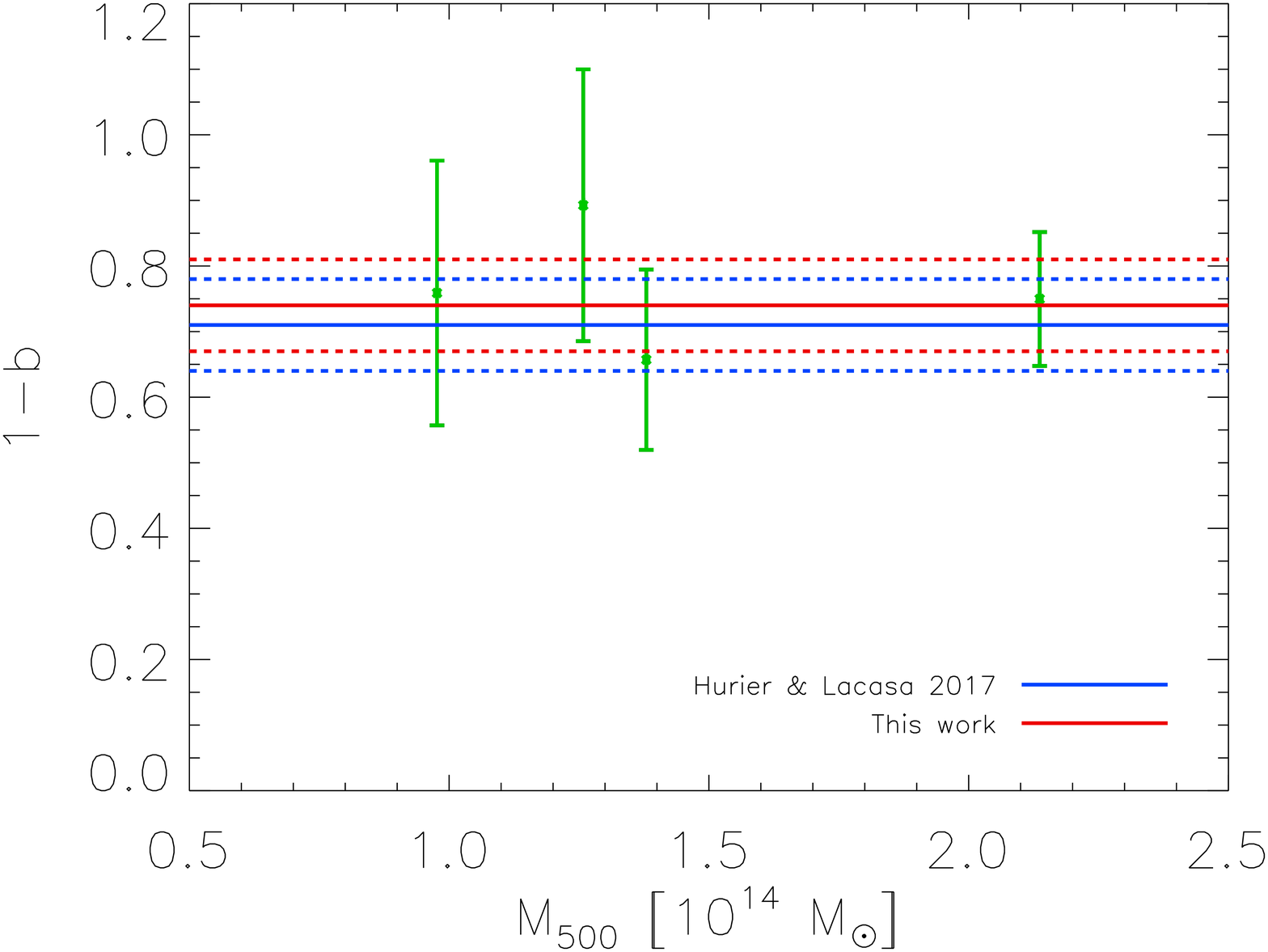}
\caption{Measurement of the hydrostatic mass bias derived from Planck tSZ and
Planck CMB weak lensing data (green symbols). The best fitting average bias is
shown as a red solid line, dashed red lines shows the 1-$\sigma$ uncertainty
level. For comparison, we also display the constraints from \citet{hur17a} as
blue lines.}
\label{bred}
\end{center}
\end{figure}

The second step in our analysis is to measure the average total tSZ integrated
Compton parameter, $Y_{\rm TOT}$, for each richness bin. We estimate $Y_{\rm
TOT}$ as the total $Y$ within an aperture of radius of 10' minus the
background, estimated in a 10' to 30' annulus. These two radii are indicated as
solid and dashed black circles in Fig.~\ref{sred}. We note that the angular
extent of clusters is expected to be smaller than 10'. Thus, our apperture
should capture the whole signal of a cluster. Due to the same reason, our
measurement should not be affected by any cluster miscentering.

We estimate the uncertainties in $Y_{\rm TOT}$ using 1000 {\tt MILCA} maps
with different realizations of the instrumental noise (correlated inhomogeneous
noise) plus CIB residuals modelled following \cite{planckszcib2}. We note that
the Planck weak-lensing map is constructed using a quadratic estimator of the
temperature and polarisation maps \citep{planckphi}. Hence, the tSZ and lensing 
maps should have uncorrelated noise and thus we neglect its impact.

We compute the hydrostatic mass bias as the ratio of the cluster mass
estimated using our tSZ procedure describe above, and the CMB weak lensing
masses presented in \citet{gea17}. We convert their mass $M_{200}$ to $M_{500}$
assuming a NFW profile \citep{nav96} and the mass-concentration relation of
\cite{duf08}. We transform $Y_{\rm TOT}$ to $Y_{500}$ assuming a
gNFW universal pressure profile \citep{arn10,planckppp}, which gives $Y_{\rm
TOT}/Y_{500} = 1.79$ \citep{planckcat}. Finally, we use the scaling relation 
$Y_{500}$-$M_{500}$ presented in \cite{planckszc}
to compute the cluster mass. This relation assumes hydrostatic equilibrium and
has been calibrated on Planck data and X-ray observations (note, however, that 
this relation does not include relativistic corrections and, consequently, in
 the previous section we used uncorrected $y$ maps).

We present the resulting values for the hydrostatic mass bias in Fig.~\ref{bred}.
Error bars include the uncertainty associated to both tSZ and CMB-lensing mass
estimates.  We derived an average bias of $b = 0.26 \pm 0.07$. This value is
consistent within 1-$\sigma$ with previous analyses performed on CMB weak
lensing \citep{mel15}, and gives slightly higher but consistent results with
previous galaxy-galaxy weak-lensing based analyses \citep[see
e.g.,][]{med17,ser17,jim17,par17,oka16,bat16,app16,smi16,hoe15,sim15,isr15,van14,don14,gru14,mah13}.
This value for $b$ is also consistent with previous results obtained from the
tSZ analysis performed in \citet{hur17a} but it favours slightly lower values than
the combined analysis of CMB and tSZ performed by \citet{sal17}. 

We fitted for an eventual mass or redshift evolution of the hydrostatic mass
bias. First we assumed the following expression,

\begin{equation}
b = b_0 + b_1 \left( \frac{M_{500} - 10^{14} [M_\odot]}{10^{15} [M_\odot]} \right).
\end{equation}

We derived, $b_0 = 0.26 \pm 0.13$ and $b_1 = -0.02 \pm 1.50$, consistent with
no mass evolution for the bias.  Then we considered,

\begin{equation}
b = b_2 \left( \frac{1+z}{1.38} \right)^\alpha.
\end{equation}

We derived, $b_2 = 0.26 \pm 0.08$ and $\alpha = 0.30 \pm 0.37$, consistent
within 1-$\sigma$ with no redshift evolution for the hydrostatic mass bias.
Given that we do not observe significant mass or redshift dependencies, we
expect our results to not be significantly affected by selection effects of the
{\it redMaPPer} cluster sample in the mass-redshift plane.

\section{Conclusion and discussion}
\label{con}

We have performed a combined analysis of tSZ and CMB weak lensing effects about
SDSS DR8 {\tt redMaPPer} clusters. The CMB weak lensing only depends on
the integrated amount of matter along the line-of-sight. Thus it offers the
opportunity to calibrate scaling relations related to baryonic physics with a
high accuracy and little to no systematic effects. 

We have inferred a value for the hydrostatic mass bias $b = 0.26 \pm 0.07$ with
no significant redshift or mass evolution. However, the significantly different
constraints derived previously from tSZ and tSZ--X-ray cross-correlation
\citep{hur15b,hur17a} may indicate that high-mass, low-z clusters tend to be
slightly less-biased. In addition, the uncertainty in our value (mostly due to
the uncertainty in CMB lensing mass estimates) prevents us from conclusive
statements about whether clusters can be modelled with a single $b$ value or
not. Nevertheless, our value for b does have implications for the tSZ-CMB
tension, as we will see next.

\begin{figure}[!h]
\begin{center}
\includegraphics[width=0.8\linewidth]{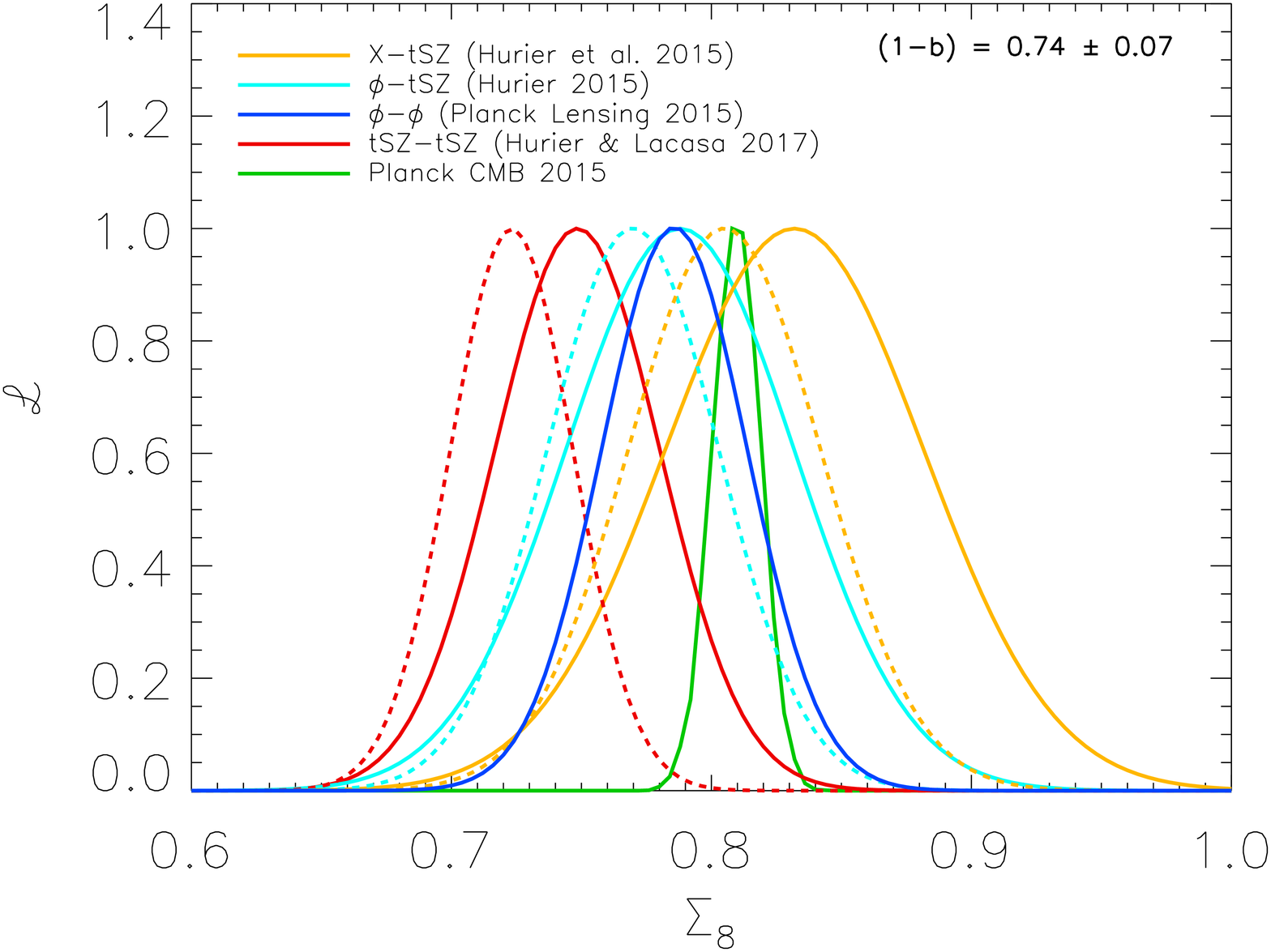}
\caption{Likelihood distribution of $\Sigma_8 = \sigma_8  \left(\Omega_{\rm
m}/0.30\right)^{0.27}$ for different analyses: tSZ angular power spectrum,
bispectrum, and number count \citep[red,][]{hur17a}, CMB weak lensing
\citep[dark blue,][]{planckphi}, CMB angular power spectrum
\citep[green,][]{planckcmb}, tSZ-weak lensing cross correlation \citep[light
blue,][]{hur15a}, and tSZ-Xray cross-correlation \citep[orange,][]{hur15b}. The
dashed lines shows the same likelihood functions assuming $b=0.2\pm0.05$
whereas the solid lines uses our measured value $b = 0.26 \pm 0.07$.} \label{lred}
\end{center}
\end{figure}

In Fig.~\ref{lred} we show several cosmological constraints for $\Sigma_8 =
\sigma_8 \ \left(\Omega_{\rm m}/0.30\right)^{0.27}$ updated by adopting our new
estimate for the hydrostatic mass bias. Specifically, we obtain: $ \Sigma_8 =
0.75 \pm 0.03$ for the tSZ analysis presented in \citet{hur17a}, $\Sigma_8  =
0.79 \pm 0.04$ for the tSZ-CMB weak lensing cross-correlation \citep{hur15a},
and $\Sigma_8 = 0.83 \pm 0.05$ for the Xray-tSZ cross-correlation
\citep{hur15b}. For comparison, we also
show the constraints assuming instead $b = 0.20 \pm 0.05$ -- a value favoured
by hydrodynamical simulations.

Our constraints bracket the expected value from CMB angular power spectrum
analysis, $\Sigma_8  = 0.81 \pm 0.01$, and are also consistent with the CMB
weak-lensing analysis (green and blue lines in Fig.~\ref{lred}, respectively).
In fact, all the cluster- and CMB-based constraints agree within $2\sigma$,
reducing the tension between the cluster abundances and the CMB. Thus suggests
that the structure in the local and early Universe are fully consistent within 
the simplest $\Lambda$CDM model.  


Conversely, our results are now in clear tension with cluster hydrodynamical
simulations that predict low values for the hydrostatic mass bias, $b < 0.2$.
This would imply the lack of important physical processes in current
simulations or large unaccounted systematic errors in observations (e.g.
selection biases in the optical cluster catalogue). In the future, more
sophisticated simulations with realistic mock observations as well as more
accurate CMB-lensing measurements will shed light on the origin of this
discrepancy.

\section*{Acknowledgement}
\thanks{We acknowledge the use of HEALPix \citep{gor05}. This project has received
funding from the Spanish Ministerio de Econom\'ia and Competitividad (MINECO)
through grant number AYA2015-66211-C2-2. REA acknowledges support of the European 
Research Council through grant number ERC-StG/716151.}

\bibliographystyle{aa}
\bibliography{szphi_red}

\end{document}